\documentclass[preprint,12pt]{elsarticle}
\usepackage{graphicx}
\usepackage{units}
\usepackage{lineno}

\newcommand{\degC}[1]{$\unit[#1]{^\circ{C}}$}
\newcommand{\mA}{\mbox{\unit{mA}}}
\def\lsim{\mathrel{\rlap{\raise 0.2ex\hbox{$\,<\,$}}{\lower 0.9ex\hbox{$\,\sim\,$}}}}
\def\gsim{\mathrel{\rlap{\raise 0.2ex\hbox{$\,>\,$}}{\lower 0.9ex\hbox{$\,\sim\,$}}}}

\journal{Nuclear Instruments and Methods A}
\begin{document}
\begin{frontmatter}

\title{
Calibration and Characterization of the IceCube Photomultiplier Tube
}

\author[Madison]{R.~Abbasi} 
\author[Gent]{Y.~Abdou} 
\author[RiverFalls]{T.~Abu-Zayyad} 
\author[Christchurch]{J.~Adams} 
\author[Madison]{J.~A.~Aguilar} 
\author[Oxford]{M.~Ahlers} 
\author[Madison]{K.~Andeen} 
\author[Wuppertal]{J.~Auffenberg} 
\author[Bartol]{X.~Bai} 
\author[Madison]{M.~Baker} 
% ...
\address[Christchurch]{Dept.~of Physics and Astronomy, University of Canterbury, Private Bag 4800, Christchurch, New Zealand}
% ...
\address[Gent]{Dept.~of Subatomic and Radiation Physics, University of Gent, B-9000 Gent, Belgium}
% ...
\address[Madison]{Dept.~of Physics, University of Wisconsin, Madison, WI 53706, USA}
% ...
\address[Bartol]{Bartol Research Institute and Department of Physics and Astronomy, University of Delaware, Newark, DE 19716, USA}
\address[Oxford]{Dept.~of Physics, University of Oxford, 1 Keble Road, Oxford OX1 3NP, UK}
\address[RiverFalls]{Dept.~of Physics, University of Wisconsin, River Falls, WI 54022, USA}
% ...
\address[Wuppertal]{Dept.~of Physics, University of Wuppertal, D-42119 Wuppertal, Germany}
% ...
\cortext[cor1]{Corresponding author}
\fntext[Erlangen]{affiliated with Universit\"at Erlangen-N\"urnberg, Physikalisches Institut, D-91058, Erlangen, Germany}
\fntext[Bari]{on leave of absence from Universit\`a di Bari and Sezione INFN, Dipartimento di Fisica, I-70126, Bari, Italy}
\fntext[Kalmar]{affiliated with School of Pure and Applied Natural Sciences, Kalmar University, S-39182 Kalmar, Sweden}
\fntext[Goddard]{NASA Goddard Space Flight Center, Greenbelt, MD 20771, USA}

\begin{abstract} 

Over 5,000 PMTs are being deployed at the South Pole to compose the IceCube 
neutrino observatory.  
Many are placed deep in the ice to detect Cherenkov light emitted by the products of high-energy 
neutrino interactions, and others are frozen into tanks on the surface to detect particles
from atmospheric cosmic ray showers.
IceCube is using the 10-inch diameter R7081-02 made by Hamamatsu Photonics.
This paper describes the laboratory characterization and calibration of these PMTs
before deployment.
PMTs were illuminated with pulses ranging from single photons to saturation level.
Parameterizations are given for the single photoelectron charge spectrum and the saturation
behavior.  Time resolution, late pulses and afterpulses are characterized.
Because the PMTs are relatively large, the cathode sensitivity uniformity was
measured.  The absolute photon detection efficiency was calibrated using 
Rayleigh-scattered photons from a nitrogen laser.
Measured characteristics are discussed in the context of their relevance to IceCube
event reconstruction and simulation efforts.
\end{abstract}


\end{frontmatter}


\section{Introduction}
\label{sec:intro}

IceCube~\cite{i3science,PDD} is a kilometer-scale high energy
neutrino telescope currently under construction at the geographic South Pole. 
A primary goal is to detect high energy neutrinos from astrophysical sources, helping to
elucidate the mechanisms for production of high energy cosmic rays~\cite{models}. 

IceCube uses the \unit[2800]{m} thick glacial ice sheet as a Cherenkov radiator 
for charged particles, for example those
created when cosmic neutrinos collide with subatomic particles in the ice or nearby rock.
Neutrino interactions can create high energy muons, electrons or tau particles, which must
be distinguished from downgoing background muons based on the pattern of light emitted.
The Cherenkov light from these particles
is detected by an embedded array of Digital Optical Modules 
(DOMs),  each of which incorporates a $10^{\prime\prime}$ diameter 
R7081-02 photomultiplier tube (PMT) made by 
Hamamatsu Photonics.  The DOMs transmit time-stamped digitized 
PMT signal waveforms to computers at the surface.

The finished array will consist of 4800 DOMs at depths of \unit[1450-2450]{m},
deployed at \unit[17]{m} intervals along 80 vertical cables, which in turn are arranged in
a triangular lattice with a horizontal spacing of approximately \unit[125]{m}.
An additional 320 DOMs will be frozen into \unit[1.8]{m} diameter ice tanks located at the surface
to form the IceTop array, which is designed for detection of cosmic
ray air showers. 
The geometrical cross sectional area will be 
$\sim$$\unit[1]{km^2}$ and the volume of ice encompassed will be $\sim$$\unit[1]{km^3}$.
Another 360 DOMs will be deployed in a more compact geometry 
(``Deep Core''~\cite{Karle2009}) using PMTs almost identical to those described here but
with a higher efficiency photocathode.

In this paper we describe measurements characterizing and calibrating IceCube PMTs,
and discuss their relevance to detector performance and event reconstruction.  
First we describe the signals of interest in Section~\ref{sec:signals}.
Section~\ref{sec:DOM} briefly describes the DOMs in which IceCube PMTs are deployed.
Section~\ref{sec:pmtgen} describes selection and basic features of the PMT, including
the dark noise rate.  
Section~\ref{sec:hv} presents the design of the HV divider circuit.
Sections~\ref{sec:freezer}--\ref{sec:absolute_calib} discuss characteristics of the
PMT in the photon counting regime, starting with single photon waveforms and charge distributions.
Time resolution is studied with a pulsed laser system.  
Uniformity of the photon detection response on the photocathode area 
is measured by scanning the entire cathode surface with a UV LED.  Absolute efficiency calibration
of the IceCube PMTs is carried out using Rayleigh-scattered light from a calibrated laser beam.
Sections~\ref{sec:linearity}--\ref{sec:afterpulse} describe response to bright pulses of light, including saturation behavior and afterpulse characteristics.

\section{Characteristics of optical signals in IceCube}
\label{sec:signals}

We begin by summarizing what the PMTs are supposed to detect, namely the optical signals
generated by neutrinos in IceCube~\cite{i3science,PDD}.
Of particular relevance are the amplitudes and widths of the pulses, requirements on time
resolution, and how the pulses are used to reconstruct physics events or reject backgrounds.
 
In detection of a high energy $\nu_\mu$ by IceCube, the neutrino interaction 
creates a muon
that traverses kilometers of ice and generates Cherenkov light along its path.
Above \unit[1]{TeV}, the muon loses energy stochastically to produce multiple showers
of secondary particles, resulting in an overall light yield proportional to
the muon energy~\cite{Lohmann,MMC}.
Most light is emitted near the Cherenkov angle, which is $41^\circ$ away from the track
direction.
The arrival times of detected photons depend on the position of each DOM 
relative to the muon's path.
For close DOMs, most photons arrive in a pulse less than \unit[50]{ns} wide, in which the
earliest photons have traveled straight from the muon track without scattering.
Significant scattering accumulates along photon 
trajectories with a characteristic length scale of about \unit[25]{m}~\cite{iceproperties},
so light pulses lengthen with distance and reach $\unit[1]{\mu s}$  (FWHM) 
for DOMs \unit[160]{m} away from a muon track.
Depending on primary energy and distance from the track, each PMT can see single photons
or pulses ranging up to thousands of photons.

Event reconstruction~\cite{i3science,PDD,dimareco} builds on principles established for the
predecessor array, AMANDA~\cite{AMA:B10,AMA:reco}.  
The observed PMT waveforms from individual DOMs are correlated and built into events,
which are fitted to physics hypotheses using the maximum likelihood method.
Each fit has access to the
complete pattern of light amplitude and timing seen by the DOMs, and
accounts for the DOMs' time response and optical sensitivity as well as time dispersion and
optical attenuation introduced by the ice.  The fit gives
the direction and energy of the muon, which in turn characterizes its parent neutrino.

The observed light pattern is also used to distinguish the rare neutrino events from the large
background of muons created in cosmic ray air showers, which are $10^6$ times more numerous.  
For this the reconstructed direction
is key, because neutrinos can come from any direction, even up from below, but the background
muons are downgoing.  A small fraction of background events 
can be misreconstructed in direction, thereby appearing to come from neutrinos, but the pattern
of detected light will generally be a poor match compared to expectations for a properly
reconstructed track.  
Misreconstruction can be aggravated by additional muons from the same shower or other 
coincident showers.
This separation between signal and background is accomplished by evaluating relative
probabilities on an event-by-event basis, and is aided by good time and amplitude resolution as well
as by low PMT noise rates.  

Similar principles apply to other types of high energy neutrino interaction.
Instead of a muon, an electron can be created that 
loses its energy in a few-meter-long particle shower~\cite{MiddellDAgostino2009}; 
on the scale of IceCube, such a shower appears almost like a point source of Cherenkov light.
For a sufficiently energetic neutrino, the light can be detected hundreds of 
meters away, and nearby DOMs can see enough light to drive their PMTs
into the nonlinear saturation regime.  
Therefore proper modeling of saturation behavior is needed
for good reconstruction and background rejection.

Design studies~\cite{PDD} for important physics goals have shown that sufficient reconstruction
quality is achieved for a
PMT timing resolution of \unit[5]{ns}, low-temperature noise rate below \unit[500]{Hz}, 
and effective dynamic range of 200 photoelectrons per \unit[15]{ns}.

In the case of lower energy (MeV) neutrinos from supernovae, IceCube
cannot resolve individual interactions.  Instead, supernovae would be detected as a 
momentary increase in the collective photon counting rate for the whole array, 
corresponding to a large number of neutrino interactions within a few seconds.  
The dark noise rate of the PMTs is particularly important here, because it dictates the statistical 
significance of any excess count rate.

IceTop uses DOMs identical to those in the deep ice.
Here the signals arise from muons, electrons and gamma rays in cosmic ray air 
showers~\cite{IceTopPDD}.
These particles deposit energy in the ice tanks housing the DOMs, resulting in light pulses up to several hundred nanoseconds long.   
The arrival times and amplitudes in the surface array are then used to reconstruct the shower core position, direction, and energy.   An overall timing resolution of \unit[10]{ns} provides pointing
accuracy of about \unit[1]{degree}.
The PMT pulses range from single photoelectrons at the periphery of showers to
\unit[$10^5$]{photoelectrons} for a \unit[1]{EeV} shower that strikes within the array.
To achieve the implied dynamic range, each tank contains two DOMs operating at gains differing by
a factor 50.

\section{The IceCube optical detector: DOM}
\label{sec:DOM}

The Digital Optical Module is the fundamental element for both optical detection 
and data acquisition in IceCube~\cite{PDD,ic22performance,i3daq}.  
It contains a $10^{\prime\prime}$ diameter PMT supported by coupling gel, the PMT high voltage
generator and divider circuits, an LED flasher board used for calibration
of the array geometry and study of ice properties, and the DOM mainboard which
contains the analog and digital signal processing electronics~\cite{i3daq}.  The PMT is
surrounded by a mu-metal grid to shield it from the terrestrial magnetic field and
improve the PMT performance.  All systems are housed within a
pressure sphere made of $0.5^{\prime\prime}$ thick glass,
capable of withstanding pressures to \unit[70]{MPa}.
The glass and gel set the short wavelength cutoff of the DOM at about \unit[350]{nm}, 
where the PMT by itself is still relatively efficient.

Strings of DOMs are deployed into water columns that have been melted by a hot-water
drill.  After refreezing, DOMs are optically well coupled to the surrounding glacial ice.
Signal and power connections between the DOMs and the surface are provided by
copper twisted-pair wires bundled together to form the main cables.  PMT signals are
digitized on the mainboard, buffered in memory, and sent to the
surface upon command of surface readout processors.  

\section{PMT selection and dark noise rate}
\label{sec:pmtgen}

A number of large-area PMTs are commercially available and have been used 
successfully to instrument large volumes in other experiments.
IceCube selected the R7081-02 made by Hamamatsu Photonics, emphasizing the criteria of
low dark noise and good time and charge resolution for single photons.
Some manufacturer's specifications are shown in Table~\ref{table:datasheet}, and
more detailed measurements are described in the following.

The nominal gain of $10^7$ was chosen to give single photon pulses around \unit[8]{mV},
which is well above the digitizer precision and other electronic noise levels (both $\sim$\unit[0.1]{mV}).
Aging was not a concern for gain selection, since
at the expected noise rates, the corresponding total charge delivered by each deep-ice PMT
will be less than \unit[1]{C} after 20 years (or \unit[100]{C} for IceTop).
Tubes with 10 and 12 stages were evaluated, with the 10 stage options showing a better
peak-to-valley ratio at this gain.
Lower gains of $5\times10^6$ and $10^5$ were chosen for IceTop PMTs because air shower
pulses generally comprise many photon detections.

\begin{table}
\caption{Hamamatsu specifications for the R7081-02 PMT (typical)~\cite{R7081Datasheet}.}
\label{table:datasheet}
\begin{center}
\begin{tabular}{lc} 
\hline
Spectral response&300 to \unit[650]{nm}\\
Quantum efficiency at \unit[390]{nm}&25\%\\
Supply voltage for gain $10^7$&\unit[1500]{V}\\
Dark rate at \degC{-40}&\unit[500]{Hz}\\
Transit time spread&\unit[3.2]{ns}\\
Peak to valley ratio for single photons&2.5\\
Pulse linearity at 2\% deviation&\unit[70]{mA}\\
\hline
\end{tabular}
\end{center}
\end{table}

The R7081-02 has 10 linear focused dynode stages and achieved the nominal gain 
of $10^7$ at about \unit[1300]{V} in our tests with the recommended divider ratios
(observed range \unit[1050-1600]{V} for 3744 PMTs). 
Its $10^{\prime\prime}$ diameter photocathode is composed of the standard bialkali material
(Sb-Rb-Cs, Sb-K-Cs) with a peak quantum efficiency of approximately 25\% at \unit[390]{nm}.
With a borosilicate glass envelope, the spectral response~\cite{PMTHandbook} is a good match 
to the spectrum of Cherenkov light after propagation through ice~\cite{iceproperties}, especially
considering the \unit[350]{nm} cutoff of the housing.

\begin{figure}[htb]
\begin{center}
\includegraphics[width=0.75\textwidth]{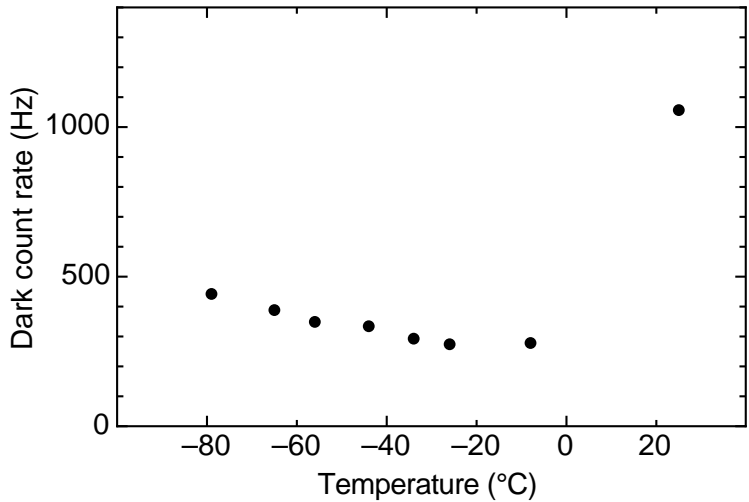}
\caption{Dark count rate versus temperature for a sample IceCube PMT covered with black tape 
(see text).
Rates were recorded after a settling time of $12$~hours of dark operation at gain $10^7$ and
discriminator threshold 0.25 times the single photoelectron peak.
An artificial deadtime circuit rejected additional hits within
 \unit[6]{$\mu$s} of each count, including about half of all afterpulses (Section~\ref{sec:afterpulse}).
The contribution from cosmic rays ($<$\unit[5]{Hz}) has not been subtracted.}
\label{fig:cvstemp}
\end{center}
\end{figure}

In response to IceCube requirements, the supplied R7081-02 units were manufactured
with a custom low radioactivity glass.
The resulting dark count rate at low temperatures is close to
\unit[300]{Hz} in the
\degC{-40} to \degC{-20} range of greatest interest for IceCube
(Fig.~\ref{fig:cvstemp}).
The higher room temperature rate can be attributed mostly to cathode thermionic emission, 
which is suppressed at low temperature.
The low temperature rate is believed to be dominated by 
radioactive decays plus scintillation in the PMT glass envelope, 
and shows a rise with decreasing temperature
similar to that reported in other studies~\cite{LowTempRise}.
The association of this rate with decays
is supported by time correlations observed on scales up to \unit[1]{msec},
as can result from
delayed particle capture or de-excitation of states created by decays.
It is also supported by the observed effect of taping: for these measurements,
the entire outside surface of the PMT glass was covered
in black vinyl tape, pulled tightly against the glass to avoid bubbles.  
The taping is observed to reduce
the low-temperature noise rate by about half. The reduction is attributed to 
absorption of outward going decay photons,
which can otherwise be channeled to the photocathode via internal reflection.
The taped result is appropriate for PMTs installed in IceCube DOMs
because they are optically coupled to gel (then glass and ice) where the
refractive index matches better than it does for air.

The low dark rate allows IceCube to record all events that satisfy simple multiplicity conditions,
and is particularly important for observation of any galactic supernova event.
Such a supernova could 
yield about $10^6$ excess photon counts in IceCube over a few seconds~\cite{Dighe:SN}.  
The single-PMT dark rate, multiplied by the number of PMTs, contributes a background
rate of $1.5\times10^6$~Hz, with a similar contribution from decays in the DOMs' glass housing.
The excess from a supernova would be easily observed above this background,
even including the details of its time structure.

On the other hand, a high energy neutrino event creates optical
pulses distributed over \unit[3]{$\mu$s}, with most information contained within a time window
less than \unit[300]{ns} wide in each DOM.
Because this window is so short, the low PMT dark rate
implies that only 1\% of muons would be accompanied by a relevant noise count
among the 100 DOMs closest to the track, and many of these DOMs detect multiple signal photons.
Therefore the degradation of reconstruction and background rejection is very small.
The dark noise rate has even less effect for IceTop DOMs, due to higher thresholds and
coincidence requirements.

\section{High voltage divider circuit}
\label{sec:hv}

\begin{figure}[htb]
\begin{center}
\includegraphics[width=\textwidth]{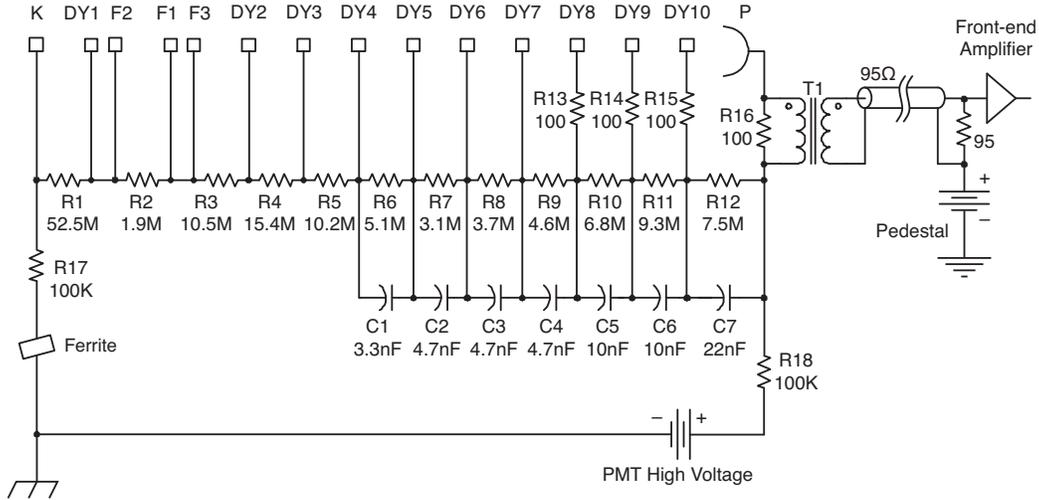}
\caption{Schematic of the passive HV divider circuit, shown with coupling to the front-end amplifier
on the digitizer board.
}
\label{fig:hvbleeder}
\end{center}
\end{figure}

The relative dynode voltage ratios for R7081-02 have been optimized by Hamamatsu
to achieve a maximum collection efficiency while achieving
$10^7$ gain between \unit[1050]{V} and \unit[1600]{V}.  
Our high voltage subsystem design fulfills the additional requirements of low
power consumption, long-term reliability, 
and sustained response to very bright pulses lasting up to a microsecond.

The dynode voltages are provided by a passive resistive divider 
with a total resistance of $\unit[130]{M\Omega}$ (Fig.~\ref{fig:hvbleeder}).
The rather high total resistance is chosen to minimize
power consumption, which is an important economic consideration for operations at the
South Pole. A custom, compact, high voltage generator~\cite{EMCO} that is both low
power ($<\unit[300]{mW}$) and low noise ($<\unit[1]{mV}$ ripple, peak-to-peak) is
used in conjunction with the passive divider. 

Capacitors are placed across the last six dynode
intervals and between the last dynode and anode.
These capacitors help sustain the PMT output for large pulses of 
up to $10^6$ photoelectrons (p.e.).
Even for illumination in the PMT
saturation region ($\sim$$\unit[200]{p.e./ns}$, see Section~\ref{sec:linearity}), the transient
gain loss after a $\unit[1]{\mu s}$ pulse ($\unit[2\times10^5]{p.e.}$) 
is observed to be less than 1\%.
A detailed simulation~\cite{photonics} 
indicates that such a pulse could arise from a \unit[50]{PeV} electron shower
\unit[100]{m} away from the PMT, which would then be faithfully recorded.
Pulses up to five times this large ($10^6$ p.e.) still result in less than 20\% transient gain loss, 
so while the primary
pulse would be completely saturated, afterpulse amplitude could be used to estimate the total
illumination (Section~\ref{sec:afterpulse}).  Finally above $2\times10^7$~p.e. the gain loss rises 
rapidly above 50\%.
For all these transient gain losses, recovery occurs within the RC time constants of order
\unit[1]{sec}.

Low-inductance resistors ($\unit[100]{\Omega}$, R13 through R15) are used to dampen ringing
that arises from coupling of the larger dynode filter capacitors with parasitic inductance 
in the dynode leads and printed circuit traces.
This ringing could otherwise be a nuisance 
when reconstructing a single PMT output waveform as a series of photon hits.

The IceCube PMTs are operated with their cathode at ground potential.
Therefore the high voltage anode is AC coupled to the front-end amplifiers.
For the AC coupling, we use a custom bifilar-wound 1:1 toroidal transformer rather than a 
DC blocking capacitor.  
High voltage reliability is achieved in the transformer winding using wire with insulation 
rated for over \unit[5]{kV} DC.
The resulting stray capacitance from anode to front-end amplifier is only
\unit[30]{pF}, which limits the stored energy which might damage the analog front-end if sparking
should occur in the HV system.
In contrast, a coupling capacitor large enough to meet the signal droop specification would be
at least 1000 times larger than the stray capacitance of the transformer.
The transformer topology also reduces noise by avoiding noisy high voltage ceramic
signal coupling capacitors and by breaking a ground loop path involving the HV power supply.
The ferrite and resistors in series with the HV supply further reduce coupling of high frequency noise
to the front end input.

The transformer coupling delivers good signal fidelity for single-photoelectron (SPE) waveforms
with risetimes of a few ns, while transmitting wide pulses
exceeding $\unit[1]{\mu s}$ with less than $10\%$ droop and undershoot.
The custom design uses 18 bifilar turns on a ferrite (Magnetics Type H) toroid core, 
providing roughly flat coupling
from \unit[8]{kHz} to over \unit[100]{MHz} at operating temperatures down to \degC{-40}.
The self-resonant frequency is above \unit[150]{MHz}.
The low operation temperature presented a challenge because the permeability 
of the transformer core
decreases rapidly with temperature, leading to a shorter droop time constant.
Although most DOMs have a time constant around $\unit[15]{\mu s}$ at ambient temperatures
near $\unit[-30]{^\circ C}$, 1200 DOMs were built using an older transformer design yielding a
time constant around $\unit[1.5]{\mu s}$ at $\unit[-30]{^\circ C}$.
The improved performance of the new design was achieved with a larger core and more turns, 
at the expense of a slightly wider SPE pulse shape (Fig.~\ref{fig:spewaveform}).
The two designs are deployed intermixed.

The droop and undershoot are relevant for the $\mu$s long trains of photon
pulses expected in DOMs over 100m away from high energy events, such as \unit[10]{TeV} electron showers or \unit[500]{TeV} muon tracks.
The small remaining effects are corrected by a software digital filter 
as a first step in event reconstruction, based on individual time constants for each DOM.
The residual error is typically less than 1\% of the pulse amplitude
(except for pulses
with peak or undershoot outside the ADC dynamic range, which is limited after 
\unit[400]{ns}~\cite{i3daq}).

The divider circuit is constructed on a \unit[10]{cm}-diameter printed circuit
board which is directly solder-mounted to the PMT. 
All components (except R13--R16) are through-hole mount type,
selected with a voltage derating factor
of two or greater (typically four) to ensure long-term reliability.
Strict voltage and voltage gradient rules are applied to the board layout. 

Coaxial cables are used for the connections to the high voltage generator
and the front end amplifier on the digitizer board.
The effective load for anode output pulses is  $\unit[50]{\Omega}$ ($\unit[43]{\Omega}$ for the older
transformer design), which includes a back-termination
resistor on the primary side of the transformer (R16), the transformer AC response, and
the input impedance of the amplifier.

\section{Single photoelectron waveform and charge}
\label{sec:freezer}

The SPE waveform shape and charge probability distribution are important for event reconstruction.
The DOM waveform digitizers are triggered when the signal reaches about 0.25 times
the typical SPE peak amplitude, after which the PMT output waveform is digitized for up to 
$\unit[6.4]{\mu s}$.  
The detection efficiency for single photons depends directly on the fraction of the SPE charge
distribution above trigger threshold.
For high energy neutrino events, many waveforms show
contributions from multiple photons, all of which could provide useful information during
event reconstruction.  The overall light yield provides an estimate of the neutrino energy, and the
space and time distribution of light helps to reconstruct direction and reject backgrounds.
The time distribution of photons can be extracted from each PMT
waveform if the response to single photons is well understood.  The response to each
photon is approximately given by the average SPE waveform, scaled randomly according 
to the complete charge probability distribution.

In order to mimic the ambient temperature in the ice, PMTs were placed in a 
freezer box at \degC{-32} and illuminated by diffused light from a 375~nm UV LED.  
The light was generated in \unit[10]{ns} pulses with intensity of about 0.1~photons per shot 
($\sim$0.02 photoelectrons per shot), dim enough to initiate only SPE signals.

\begin{figure}[hbt]
\vspace{3pt}
\begin{center}
\includegraphics[width=0.5\textwidth,clip=true]{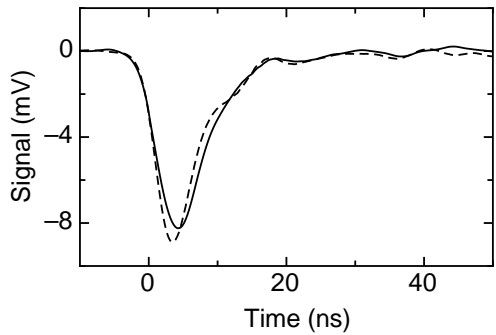}
\end{center}
\caption{Average of 10,000 SPE waveforms for one PMT at gain $1\times10^7$, as seen
at the secondary of the AC coupling transformer.  Results for other PMT samples are very similar.  The
solid and dashed curves correspond to new and old transformer designs discussed in 
Section~\ref{sec:hv}.}
\label{fig:spewaveform}
\end{figure}

Figure~\ref{fig:spewaveform} shows the average SPE waveform, measured
at the output of the AC coupling transformer with a digital storage oscilloscope (LeCroy LT374,
\unit[500]{MHz} bandwidth, \unit[0.5]{ns} samples).
Here the \unit[95]{$\Omega$} input impedance of the DOM's front end amplifier was
replaced by the series combination of a \unit[50]{$\Omega$} resistor and the oscilloscope input.

Individual waveforms have different amplitudes but their shapes are similar to
within a few percent. The waveform is dominated by a peak of Gaussian shape
($\sigma=3.2\,\rm ns$)
which accounts for 83\% of the area. 
A tail on the late side of the peak accounts for the remaining area
and exhibits a small amount of ringing.  
About 90\% of the
charge is collected before \unit[10]{ns} after the peak.
A substantial part of the observed pulse width is attributed to the damping resistors and
the coupling transformer (Section~\ref{sec:hv}).

\begin{figure}[htb]
\vspace{3pt}
\begin{center}
\includegraphics[width=0.85\textwidth,clip=true]{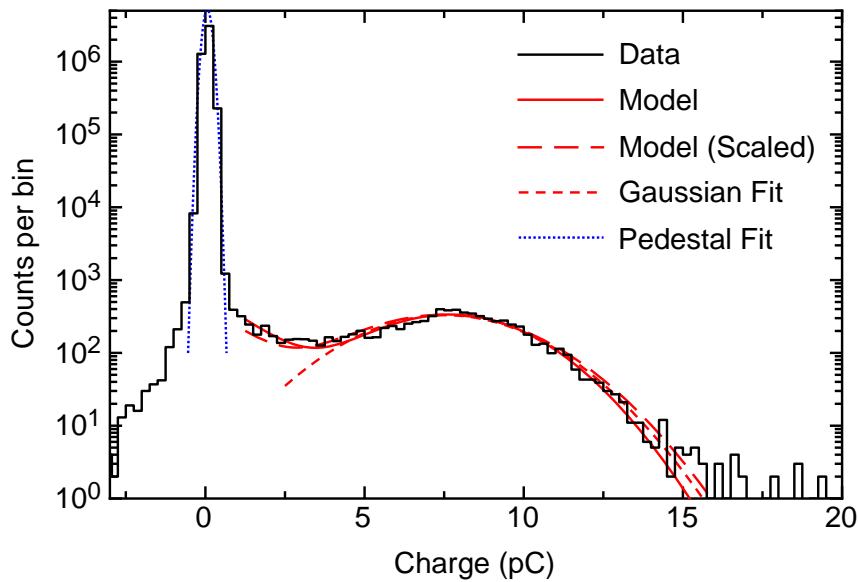}
\end{center}
\caption{Typical pedestal-subtracted 
SPE charge histogram at gain $5\times10^7$, including pedestal peak. 
Fits refer to Eq.~(\ref{eq:spe_response}), with the constraint $q_\tau/q_0=0.2$.
The remaining parameters are optimized to fit this histogram for the curve labeled ``Model'',
while ``Model (Scaled)'' optimizes only the scale parameter $q_0$ while holding 
$\sigma_q/q_0$ and $P_e$ at
values that describe the average of 120 PMTs.
}
\label{fig:SPE_fit}
\end{figure}

To study the total charge in SPE events, a computer-controlled integrating
ADC module (LeCroy 2249A) was used to integrate charge in a \unit[70]{ns} window,
triggered by the synchronization signal of the LED pulse generator.
Figure~\ref{fig:SPE_fit} shows a typical charge histogram,  which exhibits
a clear SPE peak to the right of the pedestal peak.  The Gaussian part of the
SPE peak corresponds to a charge resolution of approximately 30\%.  

The non-Gaussian component rising below 0.3 times the SPE charge in Fig.~\ref{fig:SPE_fit}
has been studied to verify that such small pulses actually reflect in-time detection of photons,
and not accidental coincidences of noise pulses such as from thermionic emission at the dynodes.
The check for a noise contribution was done with the LED light output disabled 
(but not the synchronization signal that triggers
acquisitions); all counts outside the narrow pedestal region were greatly suppressed compared to 
Fig.~\ref{fig:SPE_fit}.

The low-charge component has been
described in the past for many PMTs~\cite{dossi2000}, and
has been attributed to a
sizable probability for backscattering of the primary photoelectron at the 
first dynode~\cite{coates-backscattering-model,smirnov2004},
leading to events where only a few secondaries are produced instead of the usual 10--20.

The shape of the low-charge component is important because
even small pulses below the DOM's trigger threshold will be recorded in events
with multiple photoelectrons.
Therefore
event reconstruction should account for the entire charge probability distribution
down to zero charge, which we model as a Gaussian plus an
exponential term~\cite{dossi2000}:
\begin{equation}
f(q) = {P_e\over{q_\tau}}\exp\left[-{q\over{q_\tau}}\right] 
+ (1-P_e){1\over\sqrt{2\pi}\sigma_q}\
\exp\left[-{(q-q_0)^2\over 2\sigma_q^2}\right]
\label{eq:spe_response}
\end{equation}
Here $P_e$ is the fraction of events in the low-charge exponential part, $q_0$
is the charge at the SPE peak which defines the PMT gain, ${\sigma}_q$ is the width of the Gaussian 
fit around the SPE peak,
and $q_\tau$ is the decay constant in the exponential component.
Figure~\ref{fig:SPE_fit} shows that this is a good model for the shape of the charge histogram 
away from the pedestal.

\begin{figure}[htb]
\vspace{3pt}
\begin{center}
\includegraphics[width=0.5\textwidth,clip=true]{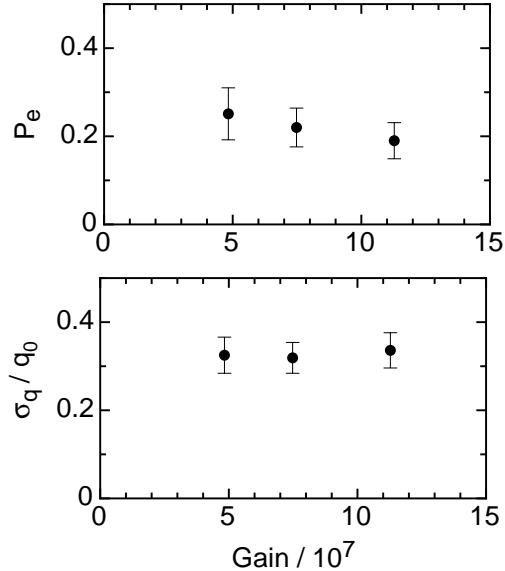}
\end{center}
\caption{
Scaled parameters from Eq.~\ref{eq:spe_response} as a function of PMT gain.
The error bars show the $1\sigma$ spread in parameters obtained for a sampling of 
115 PMTs.
}
\label{fig:model_parameters_vs_gain}
\end{figure}

Figure~\ref{fig:model_parameters_vs_gain} shows results of fitting Eq.~\ref{eq:spe_response}
for a large sample of PMTs at different gains above $5\times10^7$, excluding
the very low charge region $q<0.15q_0$ and the very high charge region
more than $2\sigma$ past the peak. 
The value of $q_\tau/q_0$ is substantially degenerate with $P_e$ for describing observed
spectra in the fitted range, so it has been fixed at the representative value of 0.20.
The scaled quantities $\sigma_q/q_0$, $q_\tau/q_0$, and $P_e$
are found not to vary strongly with the PMT gain.
The very small pulses with $q<0.15q_0$ were omitted to avoid confusion with the tail of the
pedestal distribution; results were the same if the low-charge cut was moved
to $0.25q_0$.  
The charge resolution $\sigma_q/q_0$ has been separately studied for gains between
$10^7$ and $10^8$ and again no significant effects were seen.

Figure~\ref{fig:model_parameters_vs_gain} also shows the spread in parameters from PMT to PMT.
The distribution in each parameter is approximately Gaussian, with the width shown by the error bars.
The spread is substantial, but is not expected to have a large
effect on data analysis, 
so the IceCube PMTs do not need to be parameterized individually.
Instead,
an average model is currently used in event simulation and reconstruction,
without modeling the spread.
The similarity among PMTs can also be gauged from 
Fig.~\ref{fig:SPE_fit}, where data from one PMT is compared with a model curve scaled 
from the average fit results for 120 PMTs.

The above measurements were performed with diffuse light and represent an average over
the photocathode surface. 
In a separate measurement at gain $10^7$, substantial differences were observed 
as a function of position;  for example, the peak-to-valley ratio
decreased to near unity close to the edge of the photocathode, compounding the effects of
gain variation (Section~\ref{sec:2D}).

\section{Time resolution}
\label{sec:timeres} 

The timing of recorded SPE waveforms, relative to the photon arrival time, was studied
at \degC{-40} using fast pulses (FWHM \unit[50]{ps}) from a Hamamatsu PLP-10 diode laser.
Pulses were optically attenuated and diffused over the PMT face,
yielding an average of 0.04 photoelectrons per shot.  
The wavelength was \unit[405]{nm}.  

\begin{figure}[htb]
\vspace{3pt}
\begin{center}
\includegraphics[width=0.75\textwidth,clip=true]{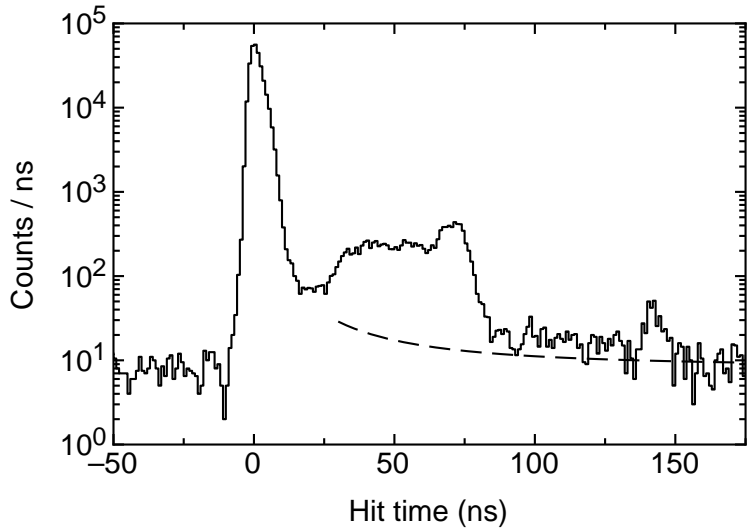}
\end{center}
\caption{Typical distribution of SPE hit times for an IceCube PMT, when illuminated
by narrow pulses from a diode laser.  
Counts are shown for \unit[1]{ns} time bins, referenced to a laser synchronization pulse.
A small fraction of late pulses (dashed line) are due to laser afterglow plus the random background
count rate; hit fractions described in the text are corrected accordingly.
The random background also explains the counts before the synchronization pulse.
}
\label{fig:time_hist}
\end{figure}

Each PMT was set for gain $10^7$ based on its SPE charge spectrum.
Hits greater than 0.4 times the SPE charge
were recorded using the DOM digitization and readout electronics.
Synchronization pulses from
the laser were also digitized to indicate the true photon arrival times to within a fixed offset.
Hit times were defined as the points where each waveform reached 50\% of its maximum, 
resulting in the time resolution histogram of  Fig.~\ref{fig:time_hist}.

The main peak of the time histogram has width equivalent to a Gaussian of $\sigma=\unit[2.0]{ns}$,
although the rising and falling edges of the peak fit better to half-Gaussians with $\sigma=1.2\rm\,ns$
and $2.7\rm\,ns$, respectively.
Some of the width can be attributed to simultaneous illumination of the entire photocathode
in our tests.
When illuminated at the center only, the width decreased to \unit[1.5]{ns};
conversely, the outer \unit[3]{cm} of the photocathode
exhibited additional delay of about \unit[3]{ns} and additional broadening.
The data acquisition system contributed a time smearing of less than \unit[0.6]{ns},
which has not been subtracted.

About 4\% of hits
are found in a shoulder (\unit[25--65]{ns}) and secondary peak at \unit[71]{ns}, and 0.2\% make up a corresponding tertiary structure (\unit[85--160]{ns}). 
The delayed hits are believed
to arise when an electron trajectory is scattered back from the first dynode
towards the photocathode, where it turns around and then eventually arrives back at the first
dynode to initiate the pulse~\cite{smirnov2004,hamamatsuPC,lubsandorzhiev}.

Because of strong photon scattering in the ice, the dispersion of hit times by the PMT 
at the \unit[2]{ns} scale is not a limiting factor for
reconstruction in IceCube;
likewise for the tail at late times.  
Considering the spacing between DOMs,
photons must typically travel tens of meters before detection, which is comparable to the scattering
length of around \unit[25]m~\cite{iceproperties}.
A detailed simulation of photon
scattering~\cite{karleSimulation} showed that at \unit[10]m distance, about 40\% of photons
are delayed by more than \unit[5]{ns}, and 10\% of photons were delayed between \unit[20]{ns} and
\unit[80]{ns}.  This is larger than the corresponding effects from the PMT itself.
On the other hand, DOMs close to a high energy track can be expected to detect at least one photon
with negligible delay, and then the very small \unit[1.2]{ns} dispersion on the early side of the 
time resolution peak may be relevant when reconstructing arrival time of the earliest photon or the
pulse rise time.

The time resolution study also reveals DOM-to-DOM differences in the nominal delay of
SPE waveforms relative to photon arrival time.
This delay includes PMT transit time plus signal delays between the PMT
output and the digitizer.  The PMT transit time is found to vary according to the square
root of the applied voltage, 
\begin{equation}
T_{transit}(V_{PMT})=T_0+2\kappa V_0\sqrt{V_0/V_{PMT}}
\label{eq:transit_time}
\end{equation}
where $\kappa=\unit[0.017]{ns/volt}$ is the slope at $V_0=\unit[1500]{V}$.
The voltage applied to each PMT is set for a design gain $10^7$, 
which is achieved between \unit[1050]{V} and \unit[1600]{V} in 99.9\% of PMTs.
The resulting RMS spread of the overall time offset is \unit[2.7]{ns}.  We find 5\% of DOMs more than
\unit[5]{ns} away from the mean, so the DOM-to-DOM corrections are currently included in
reconstruction.

\section{Two-dimensional photocathode scan}
\label{sec:2D} 

\begin{figure}
\vspace{3pt}
\begin{center}
\includegraphics[width=1.0\textwidth,clip=true]{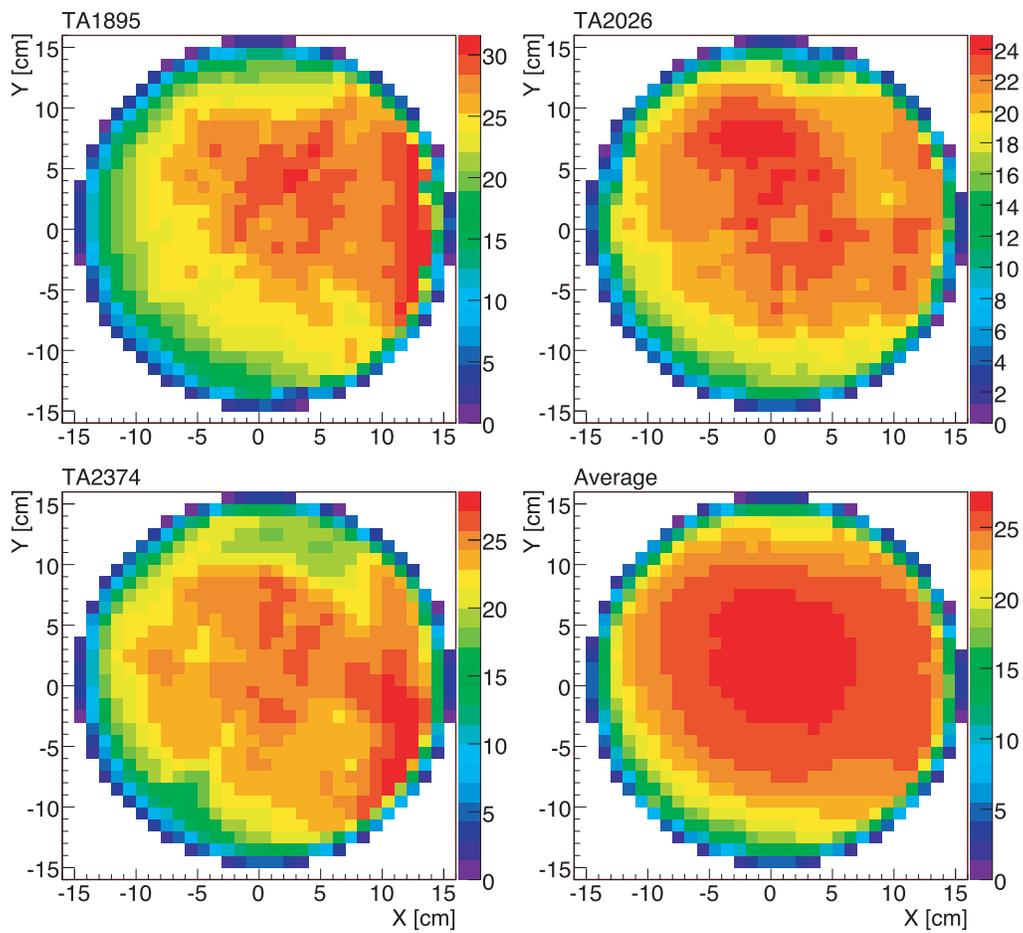}
\end{center}
\caption{Position dependence of the light pulse response for three example PMTs, and the average
of 135 PMTs (lower right).
The X-Y coordinates measure distance from the center of the photocathode
along the curved PMT face.
The color scale indicates the PMT output pulse charge in units of the SPE charge,
averaged over many pulses.
}
\label{fig:2d_CE_map}
\end{figure}

The number of photons arriving at the PMT is calculated from the observed photoelectron
signals via the PMT optical efficiency.  This is separated into an overall ``absolute
efficiency" and an angular dependence.  The dominant factor in angular dependence is
just the amount of photocathode area which can be seen from various directions.
However this has to be adjusted for the fact that the photocathode surface is very large
and different portions do not all yield the same efficiency.

We have systematically analyzed the variation of efficiency with photocathode position
at \degC{25}, using a two-dimensional scan system. 
A UV LED (\unit[370]{nm}) with collimator produced a \unit[1]{mm} spot which
was moved along the curved PMT surface, maintaining normal incidence of the light.
The LED delivered approximately
125 photons per \unit[80]{ns} pulse.  
The anode voltage was set for gain $10^7$  at the center of the photocathode,
as measured by the SPE charge peak 
($q_0$ in Eq.~\ref{eq:spe_response}).
The PMT pulse charge for each position was then measured 
by an integrating ADC triggered by the LED pulser.

Figure~\ref{fig:2d_CE_map} shows typical response maps on the cathode surface.
The measured charge is proportional to the net photomultiplier efficiency and
reflects the combined position dependence of photocathode quantum
efficiency, collection efficiency, and dynode multiplication.
PMT to PMT variation of the efficiency at a given spot on the photocathode
may be as great as 40\%, however, the spread in the area-integrated 
efficiency from PMT to PMT is much smaller, of order 10\% 
(see Sec. \ref{sec:absolute_calib}).  
The average map shows a uniform falloff in the edge region, except for a 
small bias in the +X direction.  All PMTs were measured in the same orientation, so this
bias could be associated to the first dynode position or the geomagnetic field.

A small part of the variation seen in the scans can be attributed to systematic errors, which 
arise mainly from the geomagnetic field and LED luminosity variance.
The geomagnetic field of \unit[462]{mG} is attenuated by about 50\%
with a shield made of $\mu$-metal sheet and wire, as also used in IceCube DOMs. 
By comparing measurements with the PMT rotated from its standard orientation 
in various ways, we determined
the field's effect on the efficiency variation is about 10\%. 
Because the magnetic shield is the same, the overall results are representative of what 
is expected for deployed IceCube DOMs.
(The field at the South Pole is \unit[553]{mG}, and more vertical relative to the PMT axis.)
The time dependence of LED luminosity affected the shape of each scan by less than 2\%, 
as seen by reproducibility of the scan results.

By reducing the light intensity to give only SPE hits, a similar map has been constructed
for gain variation.  
The gain can vary as a function of position because the corresponding
photoelectron trajectories arrive differently at the first dynode, leading to different yields
of secondary electrons.
The observed gain (defined by $q_0$ in Eq.~\ref{eq:spe_response})
varies within $\pm 10\%$ over the active region when high voltage is
set for gain $5\times10^7$ at the center.
However, the low-charge contribution to the SPE charge spectrum 
($P_e$ in Eq.~\ref{eq:spe_response}) was found to also vary with position, 
so that the peak-to-valley ratio decreases 
close to unity near the edge of the photocathode.
In this way, the average charge delivered per photoelectron was observed 
to decrease by up to 30\% at nominal gain $10^7$.
Because of these effects, the detection efficiency map for single photons using a specific 
discriminator threshold can differ somewhat from
the maps of Fig.~\ref{fig:2d_CE_map}.

The integrated sensitivity for a broad beam of photons incident from a particular direction
follows from the cathode efficiency maps by
averaging over the surface seen from that direction~\cite{romeo}, with correction for
non-normal incidence on surface elements as appropriate~\cite{motta2005}.
For this purpose the relative efficiency map is assumed not to vary with wavelength, {\it i.e.},
each position is assumed to obey the spectral response curve given by the 
manufacturer~\cite{PMTHandbook}.
The averaging substantially
reduces the effect of variations over the surface so the sensitivity is not strongly
dependent on direction at moderate polar angles.
Only 
light that arrives at large polar angles relative to the PMT axis will primarily
illuminate the equator region and therefore show strong azimuthal dependence.
Likewise, the variation in charge spectrum from center to edge has little effect after averaging.

The scans were performed on a small fraction of the IceCube PMTs, so only the average
variation with polar angle is used in simulation and reconstruction.  
For IceCube, the remaining PMT-to-PMT variation in directional 
sensitivity has very small consequences, because light is
typically scattered after traveling about \unit[25]{m} through the ice, and additional scattering
takes place in the refrozen ice in the hole where the DOMs are deployed.
The PMT-to-PMT variation, as well as the position dependence itself, could be more important for
detectors deployed in water where scattering lengths are much longer~\cite{water-detectors}.

\section{Absolute efficiency calibration}
\label{sec:absolute_calib} 

The absolute calibration of PMT optical efficiency is important because IceCube uses the
observed number of photons to estimate energy in reconstructed neutrino interactions.  
Showers initiated by electrons or tau leptons yield light in proportion to the energy,
and so do muons above \unit[1]{TeV} where energy loss is dominated by 
direct pair production, photonuclear interactions and bremsstrahlung~\cite{Lohmann,MMC}.

Optical efficiency can be studied after deployment by using light from muons
(produced in cosmic ray showers above IceCube) or from calibrated beacons deployed
in the ice nearby.  However it is hard to isolate the PMT response from the effects of light 
scattering and attenuation in the ice, which have some uncertainties~\cite{iceproperties}.  

Here we describe the laboratory calibration of standard PMTs installed 
in 16 IceCube DOMs distributed throughout the array.  
The calibrated PMTs provide direct information for
the energy calibration of IceCube, and will also help clarify the ice effects in other studies,
which is an important subject on its own.

\subsection{Technique}
\label{sec:abs_technique} 
\begin{figure}
\vspace{3pt}
\begin{center}
\includegraphics[width=0.9\textwidth,clip=true]{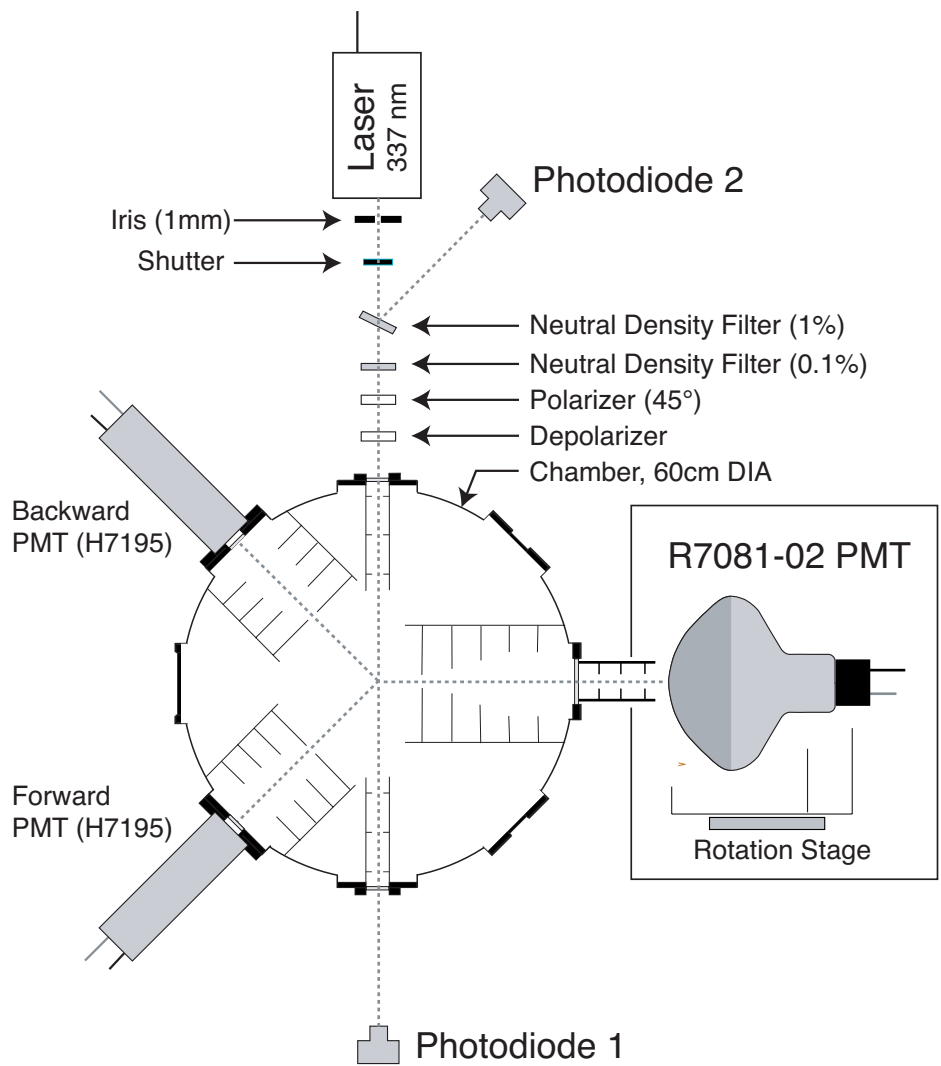}
\end{center}
\caption{Schematic view of the absolute calibration system.  Photodiode 1 establishes
the beam intensity, which is used to predict the amount of Rayleigh-scattered light reaching 
the R7081-02 PMT.  The Hamamatsu H7195 PMTs and the other photodiode are for monitoring.
Optical baffles define the scattering geometry, and chamber windows are anti-reflection coated
UV quartz ($R<0.1\%$ at \unit[337]{nm}).  }
\label{fig:QE_setup}
\end{figure}

Our setup for measuring a PMT's UV photon detection efficiency is shown in 
Fig.~\ref{fig:QE_setup}.
A pulsed 337~nm laser beam is passed through a chamber containing pure nitrogen gas, and 
the PMT to be calibrated is illuminated by the tiny amount of light that is Rayleigh scattered at about 90 degrees from the beam.  The PMT is rotated inside the dark box to probe different positions on the photocathode surface.
The primary beam intensity is measured with a calibrated silicon photodiode ``energy probe",  
and sets the fundamental scale for our efficiency measurement.  
A pressure gauge and temperature sensor provide the gas density.  
Beam intensity, geometry and gas density are then folded with the well-known Rayleigh scattering 
cross section to obtain the absolute number of photons per pulse incident on the PMT.  
Individual photon detections are counted in each pulse and divided by the number incident to
obtain the optical sensitivity at \unit[337]{nm}.  Additional corrections (Sec.~\ref{sec:abs_corrections})
are needed to obtain DOM efficiency at wavelengths around \unit[400]{nm} where IceCube is most
sensitive.

The measurement combines effects of quantum efficiency and collection efficiency, 
and can be directly applied in IceCube analysis.
It is different from the usual quantum efficiency measurement, which is based on cathode current response to a calibrated DC light source~\cite{PMTHandbook,Mirzoyan}.

The laser (Spectra-Physics VSL-337ND-S) emits \unit[4]{ns} pulses containing $\sim$$10^{10}$
photons, as measured by the silicon photodiode probe 
(Laser Probe, Inc., RjP-465).
After a warm-up delay, pulse energies are stable to within $\pm2\%$.

The beam width is \unit[1]{mm}, so the illuminated gas volume may be considered as a line source of Rayleigh scattered light. 
Apertures between the beam and the PMT define a source region with effective length 
of about \unit[1]{cm} and a spot size on the PMT of about \unit[1.5]{cm}.  
Photons can reach the PMT if they are scattered from the source region into a solid angle of 
about $\unit[7.6\times 10^{-4}]{sr}$ 
around the
$90^{\circ}$ direction, while other scattered photons are eventually absorbed on baffles or other 
surfaces inside the chamber.

The Rayleigh scattering cross section for a circularly polarized beam on nitrogen gas is taken as~\cite{RSCrossSection}:
\begin{equation}
{d\sigma_{R}\over d\Omega}=
{3\over16\pi}(1+\cos^2\theta)\times(3.50\pm0.02)\times\unit[10^{-26}]{cm^2}
\label{eq:RS}
\end{equation}
with $\theta$ as the polar angle relative to the beam direction.
The geometrical integration over the source region and corresponding solid angles
is handled in a detailed ray-tracing calculation.  After accounting for pressure and
temperature, this yields
the overall number of scattered photons reaching the PMT, typically 0.5 per pulse.
With detection efficiency around 20\%, this corresponds to
$\sim$\unit[0.1]{SPE} per pulse.

For counting photon detections,
the PMT output charge is integrated for each laser pulse with a
CAMAC ADC.  The gating time is
\unit[184]{ns} which is long enough to include the late PMT pulses described in
Section~\ref{sec:timeres}.  
The PMT gain is set close to $10^8$ as defined by the SPE peak, $q_0$ in 
Eq.~\ref{eq:spe_response}.
We then count the number of events with charge $q$ greater than a threshold $q_\mathrm{th}=0.5q_0$,
which can be clearly discriminated in the charge histogram. 
A small correction for events with multiple photoelectrons yields the number of detected
photons with $q>q_\mathrm{th}$.  

The PMT efficiency $\eta$ for $q_\mathrm{th}=0.5q_0$ is then given by comparing the number of
detected photons to the number reaching the PMT.
The efficiency for other charge thresholds can be computed by extrapolation with
the SPE charge response model, Eq.~\ref{eq:spe_response}.

\subsection{Results}
\label{sec:abs_results} 

\begin{figure}
\vspace{3pt}
\begin{center}
\includegraphics[width=1.0\textwidth,clip=true]{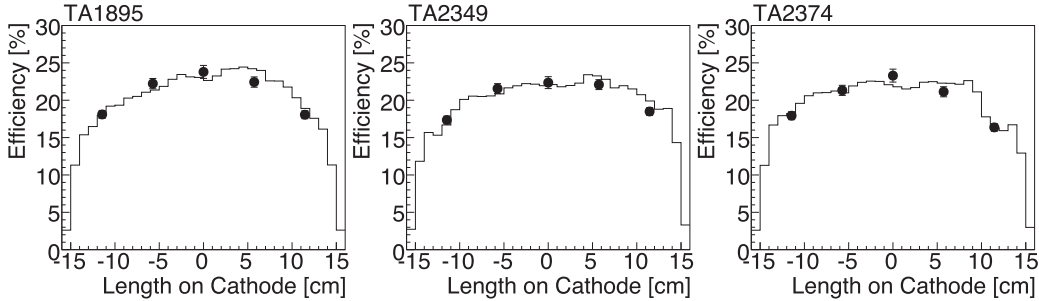}
\end{center}
\caption{Detection efficiency as a function of distance from the cathode center
for three different PMTs.  The points with error bars show absolute efficiency
measurements extrapolated to $q_{th}=0$.
The histogram curves show corresponding results from the
2D relative efficiency scans (Section~\ref{sec:2D}) after normalization.
Systematic uncertainties for the histograms are described in Section~\ref{sec:2D}.}
\label{fig:absolute_calib_qe}
\end{figure}

Figure~\ref{fig:absolute_calib_qe} shows the measured detection efficiency
as a function of distance from the cathode center.
It also shows that the absolute efficiency measurements
follow closely the shape expected from the 2D relative efficiency scans (Section~\ref{sec:2D}).
Consequently, the relative efficiency scans can be normalized to the absolute measurements
and used to estimate the absolute efficiency averaged over any given area of the photocathode 
surface.

Table~\ref{table:qe_summary} lists measured efficiencies at the center and
averaged over the whole photocathode area.  The latter is defined to include
all points within \unit[15]{cm} of the PMT axis, measured along the curved surface.
Table~\ref{table:qe_summary} also includes the photon ``effective area" which means the
amount of ideal surface with 100\% efficiency that corresponds to the actual convolution of area
and efficiency.  Here it is quoted for light uniformly spread over the PMT surface, 
with normal incidence.
For use in IceCube analysis, a similar calculation is performed for unidirectional beams as
a function of the beam angle, folding in the variation of cathode response and optical effects
at material boundaries.

\begin{table}
\caption{Measured photon detection efficiency ($\eta$) and photon effective area ($A_\mathrm{eff}$)
at \degC{25} for four different PMTs at wavelength \unit[337]{nm} and gain $10^8$.  
Values for $q_\mathrm{th}=0$ were extrapolated using Eq.~\ref{eq:spe_response},
where model parameters were fit independently for each PMT.}
\label{table:qe_summary}
\begin{center}
\begin{tabular}{@{}llllll}
\hline
PMT& $\eta_\mathrm{center}$ & 
$\eta_\mathrm{whole}$ &
$\eta_\mathrm{whole}$ &
$A_\mathrm{eff} (\mathrm{cm}^2)$ &
$A_\mathrm{eff} (\mathrm{cm}^2)$ \\
& 
$(q_\mathrm{th}=0.5 q_0)$ &
$(q_\mathrm{th}=0.5 q_0)$ &
$(q_\mathrm{th}= 0)$ &
$(q_\mathrm{th}=0.5 q_0)$ &
$(q_\mathrm{th}= 0)$ \\ \hline
TA1895 & 16.4\% & 13.2\% & 18.6\% & 84 & 119\\
TA2086 & 16.5\% & 13.6\% & 18.8\% & 87 & 120\\
TA2349 & 15.1\% & 12.1\% & 17.6\% & 77 & 112\\
TA2374 & 16.4\% & 13.0\% & 17.8\% & 83 & 114\\
\hline
\end{tabular}
\end{center}
\end{table}

The detection efficiencies in the central area of the photocathode are close to
20\% if extrapolated to $q_{th}=0$.  
These values have been compared on a PMT-by-PMT basis with measurements
by Hamamatsu using cathode response to DC light sources.
We find very good agreement, which implies that the collection efficiency is
not much less than 100\% at the cathode center.

IceCube PMTs operate at lower voltages than used in this measurement (gain $10^7$ instead
of $10^8$), so collection efficiency is expected to be slightly lower.
However, this effect is expected to be concentrated at the edges where the electron optics
are less ideal, and the efficiency falloff near the edges
is obtained from the 2D relative efficiency scans (Section~\ref{sec:2D}).  Since those scans
were done at gain $10^7$, no additional correction is necessary for the results in Table~\ref{table:qe_summary}.

\subsection{Uncertainties}
\label{sec:abs_uncertainty} 
The overall systematic uncertainty $\Delta\eta/\eta$
of the PMT detection efficiency measurement is 7.7\%, as detailed in Table~\ref{table:error_budget}.  
In addition, the measurement of each position on the PMT face
has a typical statistical uncertainty of about 5\%, set by the number of SPE hits recorded.
This is reduced to about 2\% when calculating efficiency for the whole PMT by 
combining information from the individual face positions, but the extrapolation relies on the
2D map (Section~\ref{sec:2D}) which has comparable uncertainties. 

The dominant contributions to systematic error
arise from the laser beam energy measurement,
the geometry of the scattering chamber, and the geomagnetic field.  
The first two enter directly into the calculation of the number of photons reaching the PMT.
The beam energy comes from the laser energy probe, which was factory calibrated to within
5\% at \unit[337]{nm}.
The aperture uncertainty of 4\% comes from geometrical survey of the chamber, which is used
in the ray tracing program.

The ambient geomagnetic field is \unit[462]{mG} and is unshielded in the current setup.
By changing the orientation of the PMT, we showed that it affects the point-to-point response map
at the 20\% level but the average over the surface varies by only 4\%.

\begin{table}
\caption{Systematic error budget for the PMT efficiency calibration.}
\label{table:error_budget}
\begin{center}
\begin{tabular}{@{}ll}
\hline
Source & $\Delta\eta/\eta$ \\\hline
Laser beam energy & 5 \% \\
Aperture  & 4 \% \\
Ambient magnetic field & 4 \% \\
Pressure  and temperature &  1 \% \\
Polarization & 1 \% \\
Rayleigh cross section & 0.5 \% \\
Dark noise / cosmic rays & 0.2 \% \\\hline
Overall & 7.7 \% \\
\hline
\end{tabular}
\end{center}
\end{table}

The Rayleigh scattering angular distribution depends on the polarization of the laser 
beam~\cite{Reckers1997},
so care was required to limit any polarization effect.
The effect can be strong because the PMT only sees a Rayleigh scattering signal
from the vertical component of polarization: the horizontal component induces a dipole moment oscillating perpendicular to the beam in the horizontal plane, which cannot emit power to 
the PMT which is in the same direction.
On the other hand the energy probe reads the total power regardless of polarization, so 
the fraction of power in the vertical direction must be under control. 
In our setup (Fig.~\ref{fig:QE_setup}), the laser beam is first linearly polarized at $45^\circ$
and then passed through a quartz $\lambda/4$ waveplate to convert it to 100\% circular
polarization.  By rotating a linear polarizer in the beam, we verified that the resulting horizontal and
vertical components are equal to within better than 1\%, which leads to a limit of 1\% for the
corresponding systematic effect on the efficiency measurement.

Several analyses were performed to show that the scattered photons are coming from
Rayleigh scattering and not other sources such as residual suspended dust.
We checked for expected scaling with gas density down to low pressure, symmetry of scattered
light seen in forward and backward monitoring PMTs (see Fig.~\ref{fig:QE_setup}), and
repeatability of measurements after long time intervals.

The main systematic uncertainties in our method could be reduced
if desired, comparable to the current statistical precision of about 2\%.
To accomplish this, one would calibrate the silicon photodiode ``energy meter" 
at the 1\% level; measure the aperture geometry more precisely; and provide
good shielding from the geomagnetic field.

\subsection{Additional corrections}
\label{sec:abs_corrections}
There are additional steps to obtain the detection efficiency of DOMs from the PMT efficiency
measurements, and these will be reported separately along with other studies on assembled
DOMs.
Corrections include attenuation of light in the 
glass pressure housing and the gel used for optical and mechanical coupling, 
wavelength dependence
in both the PMT sensitivity (quoted by the manufacturer~\cite{PMTHandbook}) 
and the attenuation factors, and the geometry of incident rays.
These effects are included in a detailed optical simulation of the DOMs~\cite{romeo,geant4} 
which will be compared to laboratory measurements on assembled DOMs.
A full detector simulation can be used to combine the absolute efficiencies at 337~nm, the
wavelength dependences, and the spectrum of light received from neutrino interactions.
Investigations of the combined effect show that
IceCube detects signal photons in a broad range centered on about \unit[400]{nm}.
These studies will be presented elsewhere.

Our measurements were at \degC{25}, and 
some temperature dependence can be expected.  The manufacturer quotes
a temperature coefficient of $-0.2\%/\degC$ for cathode sensitivity~\cite{PMTHandbook}.
This is being directly addressed by relative measurements of optical efficiency
of assembled DOMs at $-45\degC$, $-20\degC$ and room temperature.

\section{PMT linearity and saturation behavior} 
\label{sec:linearity}

\begin{figure}
\vspace{3pt} 
\begin{center} 
\includegraphics[width=0.8\textwidth,clip=true]
{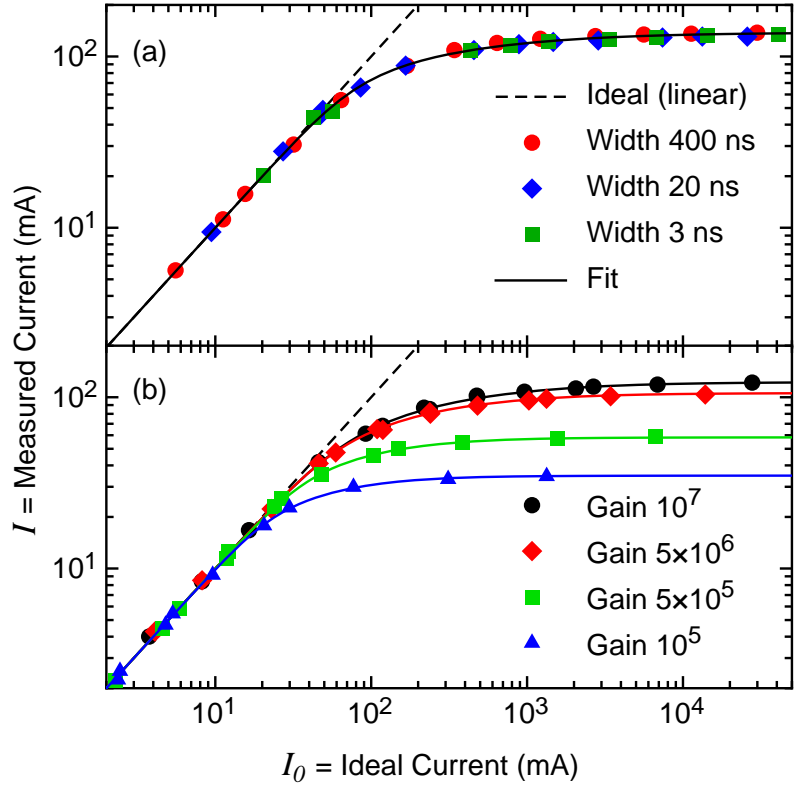} 
\end{center}
\caption{
(a) PMT saturation curve for gain $10^7$.  
The measured current is plotted against
the instantaneous light level, defined in terms of the current that would be expected for an ideal
(linear) device.  
Data points correspond to peak currents for the indicated pulse widths.
The fit curve is given by Eq.~\ref{eq:saturation_equation}, with parameters chosen optimally
for this PMT (serial number SA2747).
(b) Effect of PMT gain on the saturation curve.  Data points correspond to peak currents for
\unit[200]{ns} pulses for PMT serial number AA0020.  
The curves show the fitted parameterization,  Eq.~\ref{eq:saturation_vs_gain}.
} 
\label{fig:saturation_curves} 
\end{figure}

For most neutrino interactions expected in IceCube, any one PMT will not detect more than
a handful of photons.  For such events, and even when up to a few hundred
are detected, the PMT output is proportional to the number of photons detected.  However, some
of the most interesting signal events would be
expected to deposit large amounts of energy within tens of meters
of individual PMTs, and then the PMT response can be less than proportional.  
Optimal reconstruction requires measuring the linearity limit and modeling
the nonlinear saturation behavior.

To study saturation behavior, PMTs were illuminated with LED pulses of
various durations and brightnesses. 
The duration of the light pulses was varied from \unit[3]{ns} to \unit[1]{$\mu$s},
as measured with a fast PIN photodiode 
(\unit[1]{ns} response time). 
A set of calibrated neutral density filters was
used to control the light level. For a given LED brightness, illumination was
first measured by the PMT signal, using a filter with sufficient attenuation to
allow linear operation of the PMT. The observed signal was converted to a 
photoelectron rate and a total number
of photoelectrons (p.e.) using the SPE charge $q_0$, determined in a separate step. Then the
illumination level was increased by using different filters, with the new number of p.e.
calculated from the ratio of filter attenuation coefficients.

Figure~\ref{fig:saturation_curves}(a) shows the observed peak anode current $I$
as a function of the ideal peak current $I_0$, defined as
the peak p.e. rate times the SPE charge.
At gain $10^7$, the PMT response is linear within 10\% up to currents of
about 50mA (\unit[31]{p.e./ns}),  but saturates completely at about 150mA.
Peak responses to different light pulse widths from \unit[3]{ns} to \unit[400]{ns}
lie along a single curve.
The \unit[3]{ns} and \unit[20]{ns} width pulses were approximately Gaussian in shape, 
so the observation of identical peak response supports a saturation model where the observed current 
is a direct function of the instantaneous illumination,
with little cumulative effect from previous illumination.  In particular the data are
inconsistent with models expressed in terms of total pulse charge, which were used in some
older versions of the IceCube simulation software.
The \unit[400]{ns} pulses were approximately rectangular in shape and the output
current mirrored this shape well even in the saturation region, again as expected for
an instantaneous current saturation model.
Even long light pulses near saturation
level show only about 5\% decline from \unit[100]{mA} after 
\unit[1]{$\mu$s}
of steady illumination.
(Note this small history effect is independent of the transient gain loss due to discharge
of the dynode capacitors, which remains below 1\% for such a pulse.)

The same saturation behavior was found to apply regardless of what part of the cathode
was illuminated, even at \degC{-30}, which indicates that photocathode surface
resistance~\cite{PMTHandbook} is not important on the relevant time scales.

The instantaneous current response is well parameterized by the following:
\begin{equation} 
\ln I_0=\ln I+C{(I/A)^B\over (1-I/A)^{1/4}} \label{eq:saturation_equation} 
\end{equation} 
The parameters $A$, $B$ and $C$ differ substantially from one PMT to another
(Table~\ref{table:pmtsatfits}), so the model should not be used to invert observed
saturated pulses unless each PMT is fully characterized.  

Figure~\ref{fig:saturation_curves}(b) shows additional measurements at a range of lower
gains down to $10^5$, relevant for IceTop DOMs.  
The model of Eq.~\ref{eq:saturation_equation} continues
to apply over the full range if the parameters are scaled approximately as powers of the gain, 
as shown by the curves which are scaled by $\gamma \equiv {\rm Gain}/10^6$:
\begin{eqnarray}
A(\gamma) & = & 285 \gamma^{0.52} /(1+\gamma^{1/4})^2 \nonumber \\
B(\gamma) & = & 13 \gamma^{0.18} /(1+\gamma^{1/4})^2 \nonumber \\
C(\gamma) & = & 0.32 \gamma^{-0.13}(1+\gamma^{1/4})^2 \label{eq:saturation_vs_gain} 
\end{eqnarray}
The given parameters apply to a single measured PMT, but similar scaling behavior
can be expected for other examples; as a first estimate one would adjust the leading coefficients
in each parameter to match measurements at a particular gain, and retain the same scaling with 
gain.
Note the good numerical behavior of the scaling equations allows them to be used also for estimates
outside the given range of gain.

\begin{table}
\caption{Saturation curve parameters for three PMT samples, as defined for 
Eq.~\ref{eq:saturation_equation}.}
\label{table:pmtsatfits}
\begin{center}
\begin{tabular}{cccc} 
\hline
PMT	Serial No.& A (\mA)&B&C\\ 
\hline
AA0020&126&2.02&2.98\\
SA2747&138&2.05&3.23\\
SA2749&138&1.82&2.67\\
\hline
\end{tabular}
\end{center}
\end{table}

\begin{figure}
\vspace{3pt}
\begin{center}
   \includegraphics[width=0.8\textwidth,clip=true]{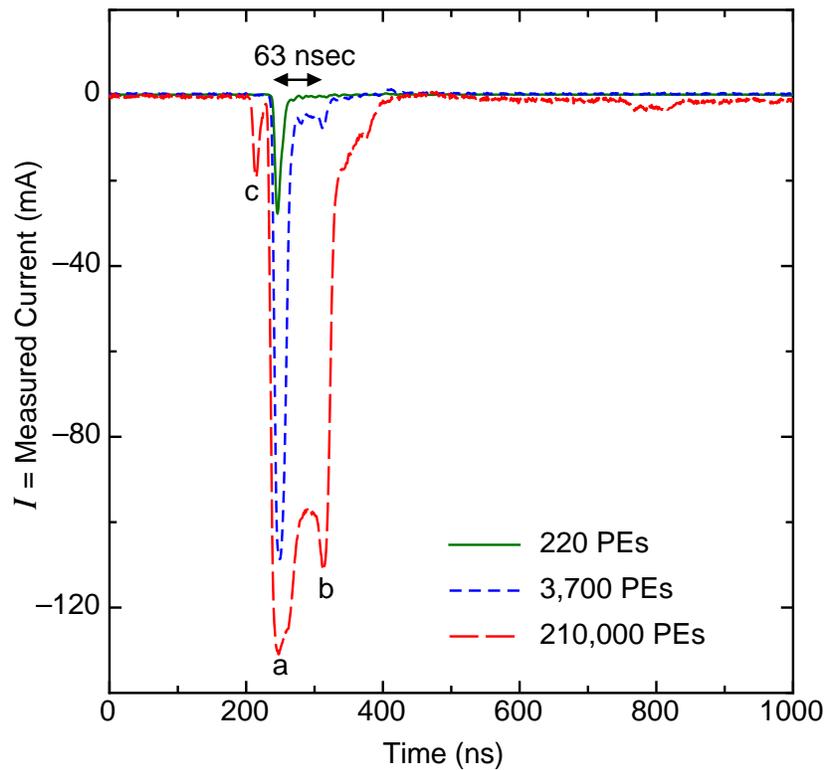}
\end{center}
\caption{Average waveforms observed in PMT serial number SA2747 
for 3~ns (FWHM) light pulses with  
progressively higher intensity: (a) main peak; 
(b) secondary peak due to unusual electron trajectories;
(c) pre-pulse.}
\label{fig:bright_narrow_pulses}
\end{figure} 
The instantaneous current model also helps understand
how the response to narrow light pulses 
(\unit[3]{ns} FWHM) broadens as intensity increases (Fig.~\ref{fig:bright_narrow_pulses}).
With the light pulse strongly
attenuated (\unit[220]{p.e.}), the PMT output pulse width is similar to the SPE
response, about 10~ns. As more light is allowed to reach the PMT
(\unit[3,700]{p.e.}), first a gradual broadening occurs to about \unit[20]{ns}
width.  
This broadening follows from Eq.~\ref{eq:saturation_equation} because the peak current
is more attenuated than the rising and trailing edges.
At this point a tail is visible, along with a second peak delayed by
about \unit[60]{ns} relative to the main peak. 
These are consistent with the late photoelectron responses
seen in SPE time resolution measurements 
(Section~\ref{sec:timeres}), except that the relative sizes of main peak and
tail are altered by saturation in the main peak. At still higher light levels
(\unit[210,000]{p.e.}), the second tail peak is comparable in size to the fully
saturated main peak, and the total width is dominated by the combination of the
two peaks. 

The highest light level in Fig.~\ref{fig:bright_narrow_pulses}
also exposes a small pre-pulse \unit[30]{ns} before
the main peak, as well as a substantial afterpulse
starting several hundred ns later (see Section~\ref{sec:afterpulse}).
The pre-pulse is ascribed to photoelectrons ejected from the first dynode,
and is somewhat exaggerated in Fig.~\ref{fig:bright_narrow_pulses} because 
the light source was aimed at
the center of the cathode with the dynode directly behind.
The individual quanta comprising the pre-pulse were separately studied using SPE-level
illumination, and were found to be between 1/10 and 1/20 of the SPE pulse size, occurring
at less than 1\% of the SPE rate.
The ratio between sizes of SPE pulses and pre-pulse quanta is similar to the typical first
dynode gain reported by the manufacturer~\cite{hamamatsuPC}.
Because individual pre-pulse quanta are below threshold for triggering DOMs, they have
a small impact on event reconstruction.  The combined pre-pulses from many photons would
only be observable for a large, narrow light pulse ($\gsim$5000 photons detected within 
\unit[30]{ns}).  Pulses originating more than $\sim$\unit[25]{m} from a DOM would generally
be broader than this, due to scattering in the ice~\cite{iceproperties}.

The saturation model (Eq.~\ref{eq:saturation_equation}) can be important for
reconstruction of very high energy neutrinos that produce electromagnetic or hadronic showers.
Ideally, reconstruction would rely 
most heavily on the PMTs closest to a shower, because the light pulse is broadened
and attenuated as it travels through the ice~\cite{iceproperties};
however these PMTs can be saturated for high energy events. 
The energy where saturation effects become important can be estimated by choosing
a characteristic distance of \unit{60}{m}, which is about half the inter-string spacing.  At this
distance, simulations~\cite{photonics} 
show that a \unit[600]{TeV} shower yields peak intensity of \unit[30]{p.e./ns},
equivalent to the linearity limit of \unit[50]{mA}.  Above this energy, signals in close PMTs
require a correction for saturation.  Above $\sim$\unit[10]{PeV}, many nearby PMTs are 
badly saturated and the shower energy measurement must rely mostly on far-away PMTs.
However, even badly saturated PMTs measure the beginning and end of the pulse,
which can be used to constrain the event geometry.

\section{Afterpulses}
\label{sec:afterpulse}

\begin{figure}
\vspace{3pt}
\begin{center}
   \includegraphics[width=0.8\textwidth,clip=true]{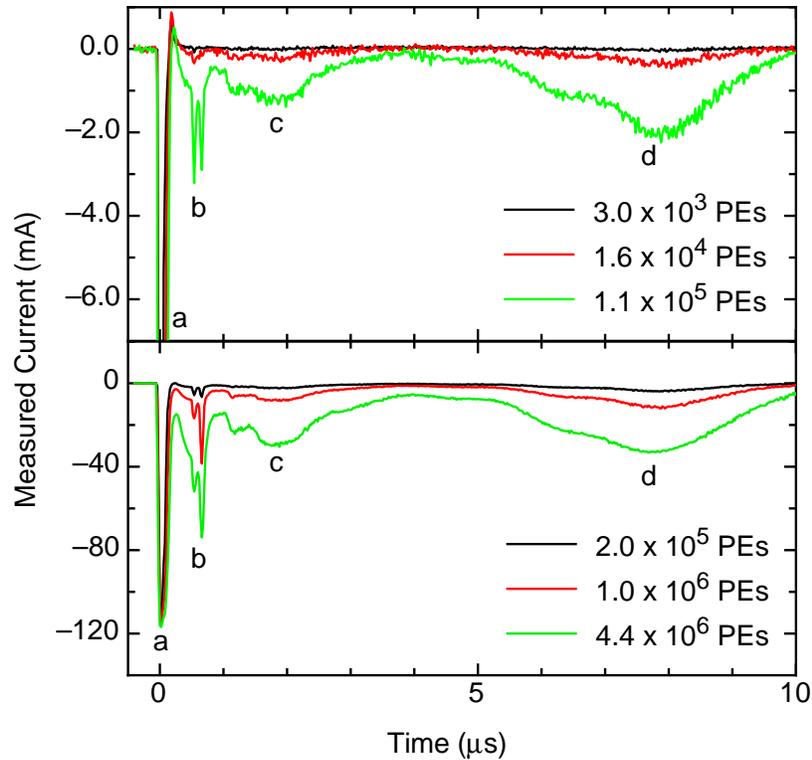}
\end{center}
\caption{
PMT afterpulse waveforms for bright flashes lasting 40~ns. Each curve is
averaged over many flashes. The primary response (a) is off-scale and saturated
at most of these intensities. Brightness of flashes is determined by the method
of Section~\ref {sec:linearity}, which is independent of PMT saturation effects.
Prominent afterpulse peaks are seen around (b) 600~ns, (c) 2~$\mu$s and (d)
8~$\mu$s after the primary response. The afterpulse waveforms grow nearly
linearly with flash intensity up to the maximum intensity measured. The PMT was
operated at gain of $10^7$, with the anode at \unit[1326]{V}. Some variation in
afterpulse waveforms is observed from one PMT to another of the same type.
Measurement of tails at very late times is slightly affected by an AC coupling time 
constant of \unit[80]{$\mu$s}.
}
\label{fig:afterpulses}
\end{figure} 
As shown in Figs.~\ref{fig:time_hist} and \ref {fig:bright_narrow_pulses}, the
prompt response to a light pulse has a tail extending to about 100~ns.
Afterpulses are seen in the range of 300~ns to 11~$\mu$s. Such afterpulses
are a common feature of PMTs, and are attributed to ionization of residual gases
by electrons accelerated in the space between dynodes~\cite{coates1973}. 
Ions created in this way can be accelerated back to the photocathode, causing ejection of electrons
which are subsequently amplified like the original photoelectrons.
Some ions strike one of the dynodes instead, but the corresponding ejecta are amplified much less
and could easily go undetected.

Afterpulse measurements were made at \degC{25} 
with LED pulses of \unit[40]{ns} width, using calibrated
optical attenuators to control the intensity, as for the saturation measurements
(Section~\ref{sec:linearity}).
In a bright LED flash, many individual ions are created, and their responses add up to an
afterpulse waveform with well defined peaks and valleys (Fig.~
\ref{fig:afterpulses}). 
The various peaks are believed to correspond to ions of
different masses, according to their individual flight times in the accelerating
field~\cite{coates1973}. Prominent afterpulse peaks for this PMT occur around
600~ns, 2~$\mu$s and 8~$\mu$s after the main response peak.  The peaks
are fairly wide and no period is entirely devoid of afterpulses until after $\unit[11]{\mu s}$.

The average afterpulse waveform grows almost linearly with the flash brightness
even up to the highest intensity studied (\unit[$4.4\times 10^6$]{p.e.} in \unit[40]{ns}), 
where the primary
response is completely saturated at $\sim$\unit[1000]{p.e./ns} (Fig.~\ref{fig:saturation_curves}). 
This suggests that observed afterpulses arise primarily from ions generated in earlier
stages of the multiplier, whose electron currents continue to rise even when later stages
have saturated.

Up to primary pulses of \unit[$1\times 10^6$]{p.e.}, the integral from
\unit[300]{ns} to \unit[11]{$\mu$s} corresponds to \unit[0.06]{SPE} per primary photoelectron.

For dimmer flashes, individual events have a small number of afterpulse
electrons. These appear as separate single afterpulses distributed in time, with probability
that can be approximately predicted from the average waveforms of Fig.~\ref{fig:afterpulses}. 
Because different ions are associated with different time ranges, and because some ions
eject multiple electrons from the photocathode, each afterpulse delay
range will be characterized by a different fundamental charge distribution.  
We have observed corresponding peak charges
from \unit[1]{SPE} to \unit[13]{SPE}, consistent with a recent more detailed study of individual
ion afterpulses~\cite{ma2009}.

The above observations are from study of only a few PMT samples, and the numbers quoted
pertain to only one (serial AA0020).  Although quantitative
differences are seen from one sample to another, the information allows one to assess
whether afterpulses affect IceCube event reconstructions, and to limit small
systematic errors.  If a particular physics analysis then appears sensitive to afterpulses, a larger
sample of PMTs would have to be studied quantitatively to provide the necessary corrections.

Given the small ratio of charge between afterpulse and primary pulse, it can be expected that
most IceCube analyses will not be strongly sensitive to the details of afterpulses.
Typical IceCube events
yield hits in each PMT that are spread over times of a few
hundred nanoseconds, well before the main part of the afterpulse distribution.
For very high energy events (e.g. electron energy \unit[1]{PeV} in the deep ice), 
signals are likely to be seen by PMTs \unit[500]{m} away where arrival times are dispersed 
over \unit[2]{$\mu$s}
(FWHM), and then the afterpulse distribution becomes more relevant.  
However, the main effect is a minor distortion of the late
part of the pulse, which already has an intrinsically long tail due to scattering.  Generally one 
does not lose much information by disregarding details of the waveforms at late times.

However, some events can have multiple peaks in the photon time distributions, and then
a characterization of afterpulses can be important for proper reconstruction.  
The most common case is an event with coincident 
arrival of one or more downgoing muons from cosmic ray showers above the detector, 
which calls for disentangling the hits originating from multiple tracks, and therefore also the
afterpulses.
Multiple muons can also arise from a single shower, and when the resulting tracks are
well separated they can yield multiple hits.
More intriguing is the possibility of $\nu_\tau$ interactions which can create two showers
of particles separated by hundreds of meters~\cite{double-bang}.  In such a case some PMTs
can see pulses of light separated by a few microseconds, so effects of afterpulses should be
considered carefully.
The late pulses described in Section~\ref{sec:timeres} should also be considered in these contexts.

\section{Summary}
\label{sec:summary} 
The R7081-02 PMT has been characterized and key findings were discussed
in the context of IceCube physics goals.  
We observe a single-photoelectron time resolution of \unit[2.0]{ns} averaged over
the face of the PMT.  A small fraction of the pulses 
arrive much later, with about 4\% between 25 and 65 nsec late. 
We also observe prepulsing and afterpulsing, with afterpulsing occurring up to \unit[11]{$\mu s$}
late.  The single photoelectron charge spectrum is well fit by a Gaussian 
corresponding to charge resolution near 30\%, 
plus a contribution at low charge which is represented by an exponential. 
The dark rate was measured to be \unit[300]{Hz} in the temperature range
\degC{-40} to \degC{-20}.
A new method for optical sensitivity calibration has been demonstrated, 
which uses Rayleigh scattering
to scale from the intensity of a primary laser beam to the much smaller number of photons 
reaching a target PMT.
Measurements of dark rate, single photon detection efficiency, single photoelectron waveform and
charge, time resolution, large pulse response, and afterpulses will serve as input for detailed 
simulation of IceCube physics events.

\section*{Acknowledgments}

We acknowledge support from the following agencies:
U.S. National Science Foundation-Office of Polar Program,
U.S. National Science Foundation-Physics Division,
University of Wisconsin Alumni Research Foundation,
U.S. Department of Energy, and National Energy Research Scientific Computing Center,
the Louisiana Optical Network Initiative (LONI) grid computing resources;
Swedish Research Council,
Swedish Polar Research Secretariat,
and Knut and Alice Wallenberg Foundation, Sweden;
German Ministry for Education and Research (BMBF),
Deutsche Forschungsgemeinschaft (DFG),
Research Department of Plasmas with Complex Interactions (Bochum), Germany;
Fund for Scientific Research (FNRS-FWO),
FWO Odysseus programme,
Flanders Institute to encourage scientific and technological research in industry (IWT),
Belgian Federal Science Policy Office (Belspo);
Marsden Fund, New Zealand;
Japan Society for the Promotion of Science (JSPS);
M.~Ribordy acknowledges the support of the SNF (Switzerland);
A.~Kappes and A.~Gro{\ss} acknowledge support by the EU Marie Curie OIF Program;
J.~P.~Rodrigues acknowledge support by the Capes Foundation, Ministry of Education of Brazil.


\end{document}